\documentclass[twocolumn,showlabels,showrefs,amsmath,amssymb,superscriptaddress,floatfix,colors]{FrontiersinVancouver}
\usepackage{lineno}
\usepackage{graphicx}
\usepackage{dcolumn}
\usepackage{bm}

\usepackage{graphicx}
\usepackage{dcolumn}
\usepackage{bm}
\usepackage{amssymb}
\usepackage{hyperref}
\usepackage{multirow}
\usepackage[english]{babel}
\usepackage{color}
\usepackage[normalem]{ulem}
\newcommand{\demian}[1]{{\color{red} #1}}

\usepackage[export]{adjustbox}

\newcommand*{\sumcirclearrowleft}{%
	\DOTSB
	\mathop{
		\mathchoice
		{\rlap{\kern.25em\rotatebox[origin=c]{-90}{$\circlearrowleft$}}{\sum}}
		{\vcenter{\rlap{\kern.2em\rotatebox[origin=c]{-90}{$\scriptscriptstyle\circlearrowleft$}}}{\sum}}
		
		{\sum}{\sum}
	}\slimits@
}

\newcommand*{\sumcirclearrowright}{%
	\DOTSB
	\mathop{
		\mathchoice
		{\rlap{\kern.25em\rotatebox[origin=c]{90}{$\circlearrowright$}}{\sum}}
		{\vcenter{\rlap{\kern.2em\rotatebox[origin=c]{90}{$\scriptscriptstyle\circlearrowright$}}}{\sum}}
		{\sum}{\sum}
	}\slimits@
}
\makeatother

\setcitestyle{square}

\def\keyFont{\fontsize{8}{11}\helveticabold }
\def\firstAuthorLast{Rouzaire and Levis} 
\def\Authors{Ylann Rouzaire\,$^{1,2,3,*}$ and Demian Levis\,$^{1,2}$ }
\def\Address{
$^{1}$Departament de F\'isica de la Materia Condensada, Universitat de Barcelona, Barcelona, Spain

$^{2}$ UBICS University of Barcelona Institute of Complex Systems , Barcelona, Spain  

$^{3}$ CECAM, Centre Europ\'een de Calcul Atomique et Mol\'eculaire,
	Ecole Polytechnique F\'ed\'erale de Lausanne (EPFL), Lausanne, Switzerland}

\def\corrAuthor{Ecole Polytechnique F\'ed\'erale de Lausanne (EPFL), Batochime, Avenue Forel 2,1015 Lausanne, Switzerland}

\def\corrEmail{rouzaire.ylann@gmail.com}

\begin{document}
	
\title[Topological Defects in the 2D Kuramoto Model]{Dynamics of Topological Defects in the noisy Kuramoto Model in two dimensions}

\onecolumn

\author[\firstAuthorLast ]{\Authors} 
\address{} 
\correspondance{} 
\extraAuth{}

	\maketitle

	\date{\today}
	
	\begin{abstract}
		We consider the two-dimensional (2D) noisy Kuramoto model of synchronization with short-range coupling and a Gaussian distribution of intrinsic frequencies, and investigate its ordering dynamics following a quench. 		We consider both underdamped (inertial) and over-damped dynamics, and show that the long-term properties of this intrinsically out-of-equilibrium system do not depend on the inertia of individual oscillators. The model does not exhibit any phase transition as its correlation length remains finite, scaling as the inverse of the standard deviation of the distribution of intrinsic frequencies. The quench dynamics proceeds via domain growth, with a characteristic length that initially follows the growth law of the 2D XY model, although is not given by the mean separation between defects. 	 Topological defects are generically free,  breaking the Berezinskii-Kosterlitz-Thouless scenario of the 2D XY model. Vortices perform a random walk reminiscent of the self-avoiding random walk, advected by the dynamic network of boundaries between synchronised domains;  featuring long-time super-diffusion, with the  anomalous exponent $\alpha=3/2$. 
				\tiny
 		\keyFont{ \section{Keywords:} Topological Defects, Synchronisation,  Active Matter, Langevin Dynamics,  Anomalous Diffusion, Out-of-Equilibrium Systems}
	\end{abstract}

\demian{This article is a draft (not accepted yet !)}

\twocolumn
	\section{Introduction}
	
Back in the XVII-th century, Huygens realised that two pendulum clocks, when sitting on the same board, start, after some transient time, to beat at the same frequency in phase or in anti-phase \cite{huygens1895letters}. Such spontaneous temporal coordination of coupled oscillatory objects is referred to as synchronization. Since then, the study of synchronization in large populations of oscillators has remained a recurrent problem across different sciences, ranging from physics to biology, computational social sciences or engineering \cite{strogatz2012sync, Pikovsky2003}. 

Much progress in the understanding of synchronisation has been achieved through the detailed analysis of simplified model systems. In this context, the Kuramoto model of phase coupled oscillators has (and still does) played a central role \cite{Kuramoto,AcebronRev}.  The original version of the model \cite{Kuramoto}, considering only all-to-all interactions, shows that global phase synchronisation can be achieved for large enough coupling, relative to the dispersion of the oscillators' intrinsic frequencies. 

Persistent oscillations, obviously need a constant energy supply, which at the microscopic scale is dissipated into the environment, which also provides a source of noise.  
Thus, to describe oscillator systems at the micro-scale, such as genetic oscillators \cite{Mondragon2011}, noise should be taken into account, as well as shorter-range interactions, leading to finite-dimensional noisy extensions of the Kuramoto model \cite{ArenasRev}. The situation in this case, which is the object of the present study, is quite more involved than the  original   Kuramoto model  and the literature far more scarce \cite{hong2015finite, lee2010vortices, RouzaireLevis}. 

The problem of synchronisation in physical sciences has more recently experienced a resurge of interest  in the context of active matter. Active matter stands for a class of soft matter systems, composed of coupled interacting agents that convert energy from their surrounding into some kind of persistent motion \cite{shankar2022topological}, such as biological units oscillating at a given rate. Such injection of energy at the level of each single constituent, drives active systems out-of-equilibrium. At the collective level, interactions between active agents result in the emergence of a variety of collective states. In particular, a broad class of active systems can spontaneously self-organise into synchronised states characterised by the coherent motion of self-propelled individuals, a phenomenon called flocking (a term borrowed from the spectacular example of the murmuration of starling birds) \cite{ginelli2016physics}. The interpretation of flocking in terms of synchronisation of active oscillators has been pointed out in a number of recent works \cite{chepizhko2010relation, chepizhko2021revisiting, levis2019activity}. 

Closer to the standard synchronisation set-up of oscillators lying on a static substrate, synchronised states have been studied in numerous active matter systems in the absence of self-propulsion. A salient example are active filament carpets, such as cilia or flagella attached on a surface. Cilia, for instance, perform a beating cycle (usually  modeled as a phase oscillator)  generating a net hydrodynamic flow (at  low Reynolds number) that  affects the motion of their neighbours \cite{golestanian2011hydrodynamic, solovev2022synchronization1,solovev2022synchronization2}.  Such filaments, though as coupled oscillators, might then synchronise, to optimise a biological function such as propelling microorganisms. 
Chiral colloidal fluids composed of spinning colloidal magnets constitute another promising novel venue to investigate synchronisation at the micro-scale \cite{Soni2019,massana2021arrested}. In these systems, an external oscillatory field drives the colloidal magnets, making them rotate around their body axis at a given frequency imposed by the field.   

Here we investigate the collective dynamics of Kuramoto oscillators, with short range coupling, in two dimensions (2D), in contact with a thermal bath and with a Gaussian distribution of intrinsic frequencies. In the absence of driving,  or intrinsic oscillations,  the system is equivalent to the 2D XY model, which exhibits a topological Berezinskii-Kosterlitz-Thouless (BKT) transition driven by the unbinding of topological defects \cite{Kosterlitz1973,Kosterlitz1974,Berezinskii1971}. As such, it provides a natural playground to study the role played by topological defects in systems of coupled oscillators. The dynamics of topological defects in out-of-equilibrium systems has been extensively investigated over the last decade in soft active systems \cite{shankar2022topological, PRLino},  where it has been found that, in many instances, defects self-propel, or super-diffuse \cite{RouzaireLevis, GiomiPierce,Bartolo2021, bowick2022symmetry}. 
A recent study of this model shows that defects become free upon self-spinning, although two regimes remain clearly distinct, a vortex-rich and a vortex-poor one \cite{RouzaireLevis}.
Here we focus our attention on the dynamics of the system following a quench, from a random initial state, where many defects proliferate, to the low temperature (noise) regime where defects are scarce. 
The coarsening dynamics following an infinitely rapid quench has been extensively studied in model statistical physics systems \cite{BrayRev}, but has received little attention in the context of synchronisation \cite{LevisPRX}. Of particular interest for us, the study of the coarsening dynamics of the 2D XY model has shown that topological defects, here vortices, diffuse, interact and annihilate, in a way that sheds light on the mechanisms underlying the non-equilibrium relaxation of the system \cite{YurkeHuse1993, rojas1999dynamical, bray2000breakdown, berthier2001nonequilibrium, JelicCugliandolo}. 
In this contribution, we aim at characterising the dynamics of the noisy Kuramoto model in 2D, to examine the random motion of vortices and discuss the fundamental differences displayed by the system as compared to its equilibrium limit. 

	We first introduce the model and its governing Langevin equation in \textbf{Section~\ref{sec:model}}. The main results are presented in \textbf{Section~\ref{sec:results}}: in Subsection~\textbf{\ref{sec:overdamped}}, we first focus on the overdamped regime and study the coarsening dynamics of the system following different quench protocols, focusing our analysis on the evolution of different correlation lengths and density of topological defects. We then study the relaxation dynamics in the underdamped regime in Subsection~\textbf{\ref{sec:underdamped}} to show that the inertia of individual spins (or oscillators) does not influence the overall large-scale dynamics of the system.  \textbf{Section~\ref{sec:discussion}}  is devoted to the defects's dynamics. We first discuss the breakdown of the Berezenskii-Kosterlitz-Thouless scenario in subsection~\textbf{\ref{sec:collapse}}, by investigating the effective interactions between defects. In subsection~\textbf{\ref{sec:defectsdynamics}}, we then describe the spontaneous creation process of vortices and the complex dynamics they exhibit. We argue that even though the trajectories of defects are very much reminiscent of genuine self-avoiding random walks, the full displacement statistics indicate that they are not strictly equivalent.

	\section{The Model} \label{sec:model}
	We consider a set of $N$ phase oscillators sitting on the nodes of a $L\times L$ square lattice with periodic boundary conditions (PBC). The evolution of their phase $\theta_i$ is described by the Kuramoto model. The governing equations of motion are given by the following set of coupled Langevin equations

	\begin{equation} \label{eq:model} 
		{m}\ddot{\theta}_i + \gamma \dot{\theta}_i =\sigma\omega_i + J \sum_{ j\in\partial_i} \sin(\theta_j - \theta_i) + \sqrt{2\gamma k_BT}\,\nu_i 
	\end{equation}
where the sum runs over the four nearest-neighbours of spin~$i$, denoted by $\partial_i$,  $\nu_i$ is a Gaussian white noise of unit variance and zero mean  and $\gamma$ the damping coefficient.  The system is thus in contact with a thermal bath at temperature $T$. The driving amplitude $\omega_i$ is drawn from a  Gaussian distribution with zero mean and unit variance. [Since Eq.~(\ref{eq:model}) is invariant under the transformation $\theta\to \theta - \Omega t$, one can choose a distribution of frequencies with zero-mean without loss of generality. ]. The amplitude $\sigma$  quantifies the dispersion of the intrinsic frequencies. 
	
	In the absence of intrinsic persistent oscillations, i.e. $\sigma = 0$, the model Eq.~(\ref{eq:model}) is equivalent to the 2D XY model with non-conserved order parameter dynamics \cite{hohenberg1977theory}. Indeed, Eq.~(\ref{eq:model}) can be rewritten as 
		\begin{equation} \label{eq:under} 
		{m}\ddot{\theta}_i  =- \gamma \dot{\theta}_i + \sigma\omega_i -\frac{\partial H}{\partial \theta_i} + \sqrt{2\gamma k_BT}\,\nu_i 
	\end{equation}
		 where 
		 \begin{equation} \label{eq:XYmodel} 
	H=-J \sum_{\langle i,j \rangle} \bold{S}_i\cdot \bold{S}_j\,,\ \  \bold{S}_i=(\cos\theta_i,\sin\theta_i)
	\end{equation}
	is just the classical XY Hamiltonian (here the sum runs over all the links of the lattice, or, equivalently, all nearest neighbours pairs). This system exhibits a BKT transition at a critical temperature $T_{KT}\approx 0.89$ \cite{littTKT,Hasenbusch2005}.
The distribution of natural frequencies introduces quench disorder in the XY model. The way it is introduced though, is fundamentally different to what is typically done in disordered systems, namely, adding disorder in the interactions or an external random field \cite{cardy1982random, le1995replica, RFXY}. The distribution of intrinsic frequencies drives the system out-of-equilibrium. Adding a term $\sigma\sum_i\theta_i\omega_i$ in the Hamiltonian $H\to H'=H+\sigma\sum_i\theta_i\omega_i$, would provide the same equation of motion Eq.~(\ref{eq:model}) when writing: ${m}\ddot{\theta}_i + \gamma \dot{\theta}_i = -\frac{\partial H}{\partial \theta_i} + \sqrt{2\gamma k_BT}\,\nu_i $. However, $H'$ is now unbounded, as a result of the constant injection of energy into the system needed in order to sustain the intrinsic oscillations. Thus, the system is intrinsically out-of-equilibrium and cannot be mapped to random field or random bond XY models. 
 
In equilibrium conditions ($\sigma = 0$), the nature of the dynamics does not affect the steady-properties of the system, as long as it fulfills detailed balance. Most  studies on the dynamics of the 2D XY model have been performed in the overdamped limit, namely
		\begin{equation}
		\gamma \dot{\theta}_i = \sigma\omega_i -\frac{\partial H}{\partial \theta_i}  + \sqrt{2\gamma k_BT}\,\nu_i  \ ,
		\label{eq:overdamped}
	\end{equation}
in its Langevin version, or, equivalently, using single-spin flip Monte Carlo dynamics \cite{JelicCugliandolo, berthier2001nonequilibrium,littTKT,Hasenbusch2005,RFXY}. 

However, in non-equilibrium conditions, the specific features of the dynamics might affect the resulting large scale behavior at long times. The dynamics of the model introduced by Kuramoto was also originally overdamped. Later on, the model was extended to include inertia \cite{komarov_synchronization_2014,gupta_kuramoto_2014,olmi_hysteretic_2014}. Here we study both the model with underdamped and overdamped dynamics following Equations~(\ref{eq:under}) and (\ref{eq:overdamped}), respectively. 

To identify the key control parameters of the present model, we rewrite the equations of motion in their dimensionless form (see Supplementary Material (SM) \cite{SM}), expressing time in units of $\gamma/J$. This leads to the following set of quantities: (i) the reduced temperature $\tilde{T}=k_BT/J$, (ii) the reduced frequency dispersion $\tilde{\sigma}=\sigma/J$ and, for inertial dynamics, (iii) the reduced inertia $\tilde{m}={m}\,J/\gamma^2$. From now on, we shall make use of these reduced parameters and drop the tilde $\tilde{T}, \tilde{\sigma}, \tilde{m}\to T,\sigma,m$.

We integrate numerically Equations~(\ref{eq:under}) and (\ref{eq:overdamped}) using a modified velocity Verlet algorithm \cite{komarov_synchronization_2014} and standard Euler-Mayurama scheme (details provided in the SM \cite{SM}). The typical system size used is $L~=~200$, if not stated otherwise.

	\section{Results} \label{sec:results}
	\subsection{Overdamped Dynamics} \label{sec:overdamped}
We first study in this section the dynamics of the model in the overdamped limit, i.e. Eq.~(\ref{eq:overdamped}), following a infinitely rapid quench from an initially disordered state where all the phases are picked from a homogeneous distribution between 0 and $2\pi$.
	
\begin{figure}[]
	\includegraphics[width=0.99\linewidth]{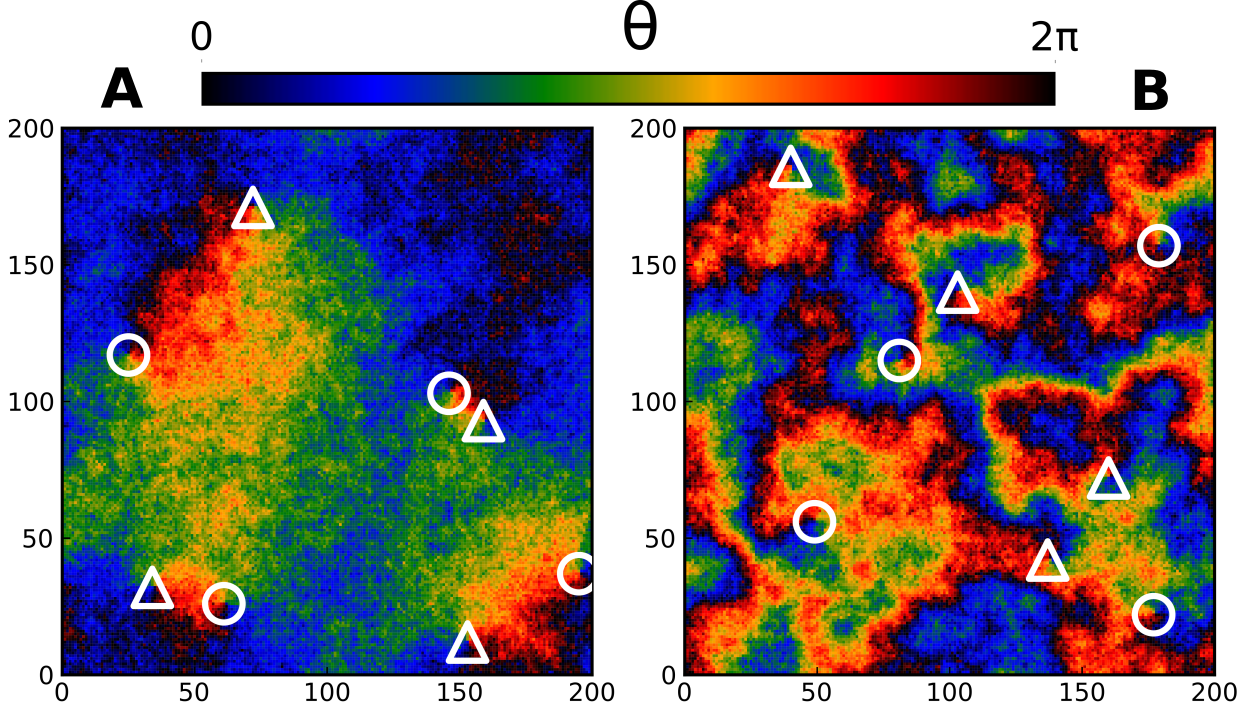}
	\caption{
		Snapshots for $T =0.2$, \textbf{(A)} $\sigma^2=0$ and \textbf{(B)} $\sigma^2=0.1$. The phase of each oscillator is represented by a color scale. Vortices (antivortices) are represented by circles (triangles).
	}\label{snapshots}
\end{figure}

In order to illustrate the nature of the low temperature state, we show in \textbf{Figure~\ref{snapshots}} representative steady state snapshots for \textbf{(A)} $T=0.2$ and $\sigma^2=0$ \textbf{(B)} $T=0.2$ and $\sigma^2=0.1$. Both snapshots contain the same number of defects: four vortices (circles) and four antivortices (triangles), visually identified as the points where black, red, yellow, green and blue regions meet. This correspond to plaquettes with a winding number $q=\pm 1$, defined by the discrete circulation of the phase differences along a plaquette, namely $\sum_{\square}\Delta\theta_{ij}=2\pi q$. 

At equilibrium, when spins are not forced, one recovers the classical XY model picture, with large regions of spins sharing the same color, ending at topological defects of opposite charge.  As the Kosterlitz-Thouless theory \cite{Kosterlitz1973,Kosterlitz1974} predicts, those defects visually come in pairs.  
From the snapshots one already observes a clear connection between the typical size of correlated domains of mostly parallel spins, and the typical distance between defects. We will come back to this point later on. 

On the contrary, upon self-spinning, the first impression is qualitatively different, even though the number of defects is identical in both panels (here 8). The phase field pattern emanating from the vortices  does not extend over large distances, and the typical size of oriented domains seem decoupled from the defects' locations, as seen in \textbf{Figure~\ref{snapshots}}\textbf{B}. Ordered regions are in this case rather localized, and one can no longer pair vortices by visual inspection of the snapshots. 
	
	\subsubsection{Coarsening Dynamics}
		\begin{figure}[]
		\includegraphics[width=0.99\linewidth]{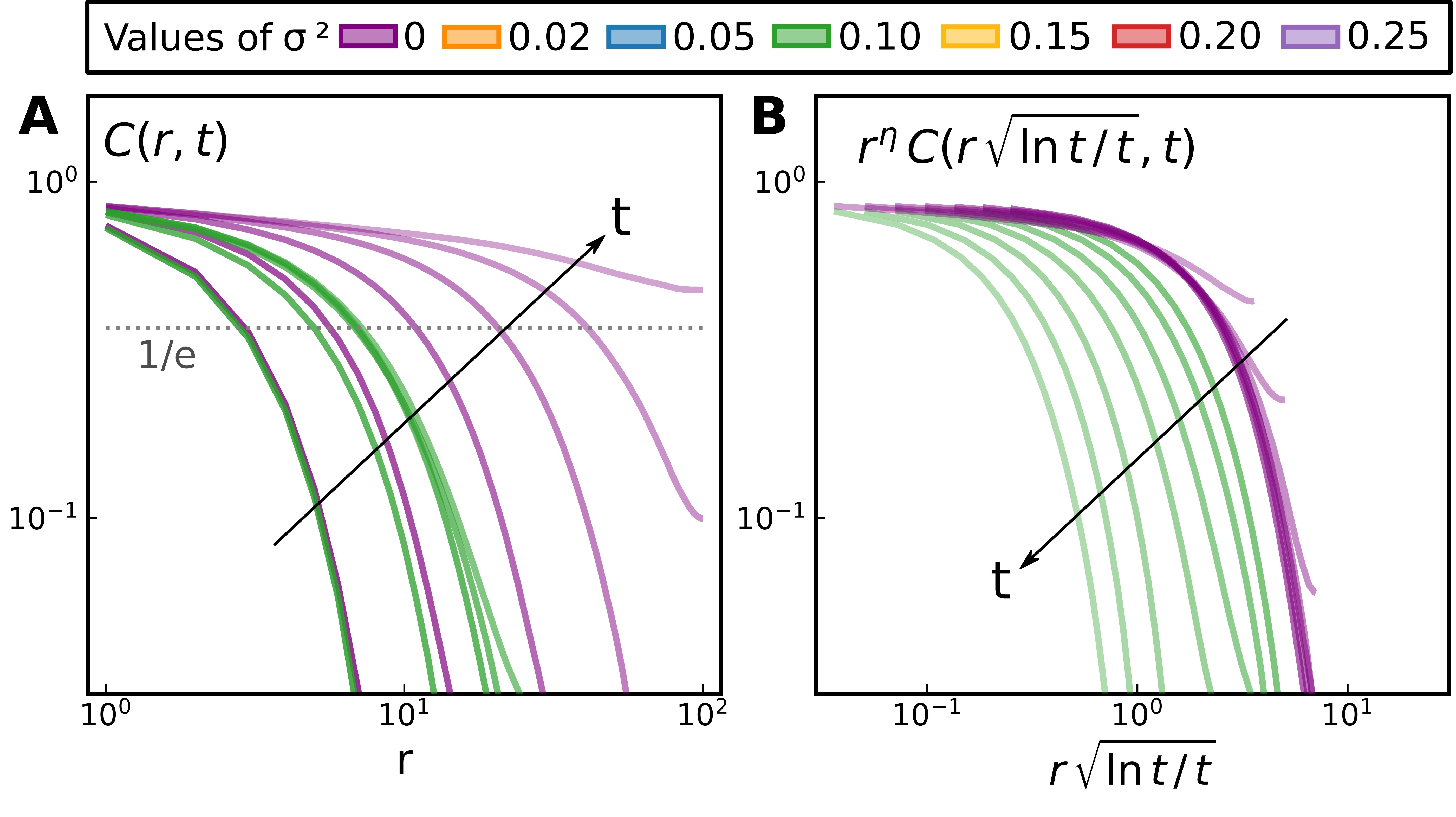}
		\caption{
		Time evolution of \textbf{(A)} the space correlation function $C(r,t)$ at different times $t\approx 4,20,100,400,2000,10000$ \textbf{(B)} the rescaled correlation function ($\eta = T/2\pi$ from the spin-wave theory) against the rescaled time. In both panels, the purple curves correspond to the XY model at $T=0.4$ and the green curves correspond to a forced system at $T=0.4$ and $\sigma^2=0.1$. Greater times are represented by lighter colors, as the arrows indicate. 
		}\label{crt}
	\end{figure}


\hspace{3cm}
\begin{figure*}[]
	\includegraphics[width=0.99\linewidth]{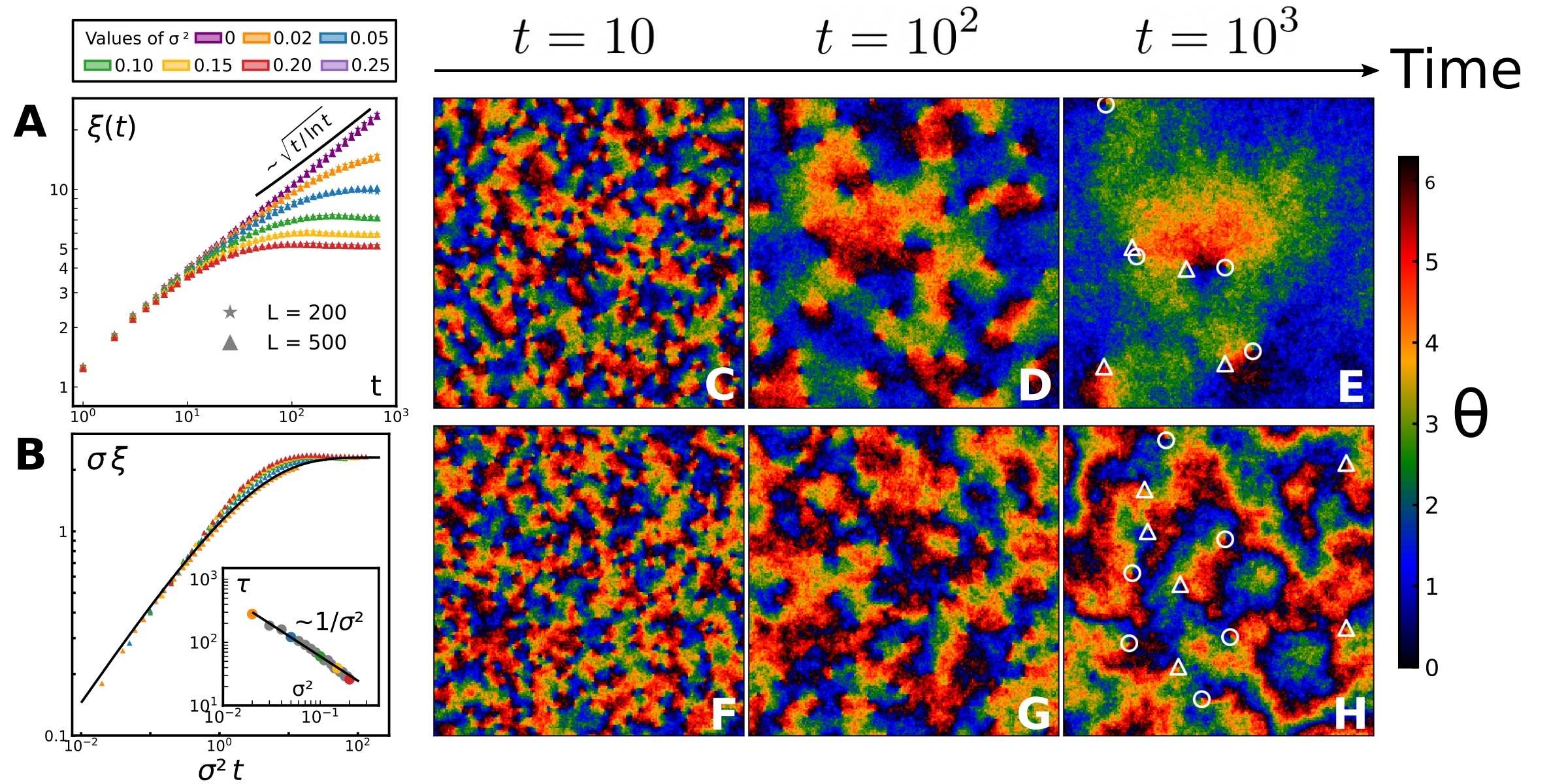}
	\caption{Time evolution of the characteristic length scale \textbf{(A-B)} illustrated by typical snapshots of a $200\times200$ system \textbf{(C-H)}.
	\textbf{(A)} Growth of the typical length scale $\xi(t)$ for  $T=0.4$ and several  $\sigma$, as indicated in the legend above (same colors as in \textbf{Figures~\ref{crt}} and \textbf{\ref{n}});   \textbf{(B)} the same data where $\xi$ and $t$ have been rescaled by $\sigma$ and  $\sigma^2$, respectively. All curves collapse onto a single master curve  $f(x) = a(1-\exp(-b\sqrt{x})$ with $a = 2.3$ and $b = 0.65$. \textbf{Inset:} variation of the transients time $\tau$ defined in Eq.~(\ref{eq:tau}) against the self-spinning intensity.
	\textbf{(C-D-E)} Representative configurations at different times showing the evolution of the system after a quench ( for $T =0.2$ and $\sigma^2=0$). The phase of each oscillator is represented by a cyclic color scale. In the last panel only, vortices (anti-vortices) are represented by circles (triangles). \textbf{(F-G-H)} Representative configurations at different times now for $T =0.2$ and $\sigma^2=0.1$. 
	}\label{xi}
\end{figure*}

	\begin{figure}[]
	\includegraphics[width=0.99\linewidth]{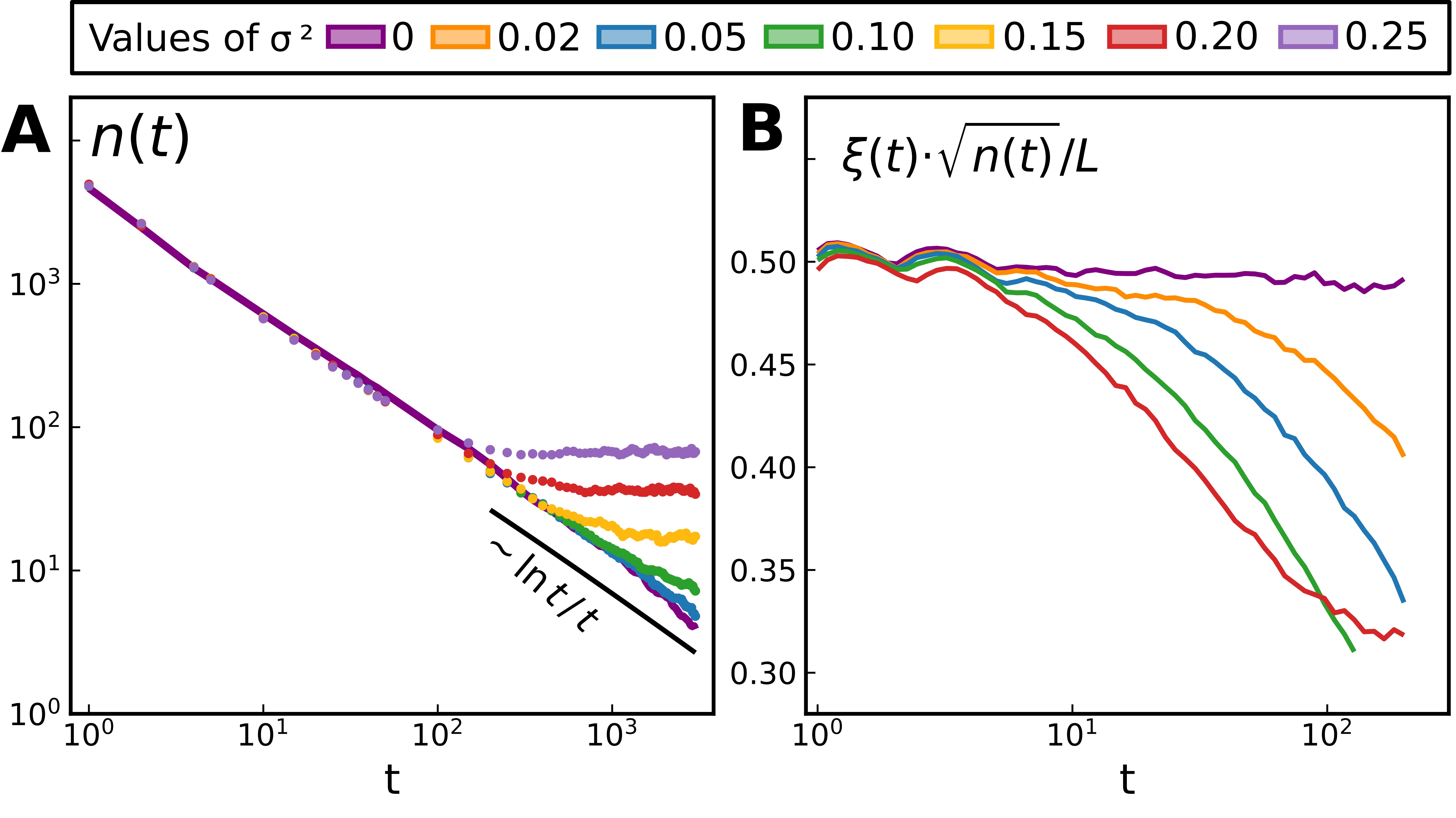}
	\caption{
		Time evolution of \textbf{(A)} the total number of defects $n$ for different forcing intensities and $T=0.4$ \textbf{(B)} the quantity $\xi\,\sqrt{\rho}$ where $\rho = n/L^2$ is the defect density. Note that because the defects are homogeneously distributed, $\sqrt{\rho}$ is the average distance between two of them. 
	}\label{n}
\end{figure}

To gain quantitative insight, we compute the spin-spin correlation function 
	\begin{equation}
	 	C(r,t)=\langle \cos(\theta(r,t) - \theta(0,t))\rangle \ ,
	 \end{equation} 
where the angular brackets denote the average over spins and independent realisations of both thermal noise and quenched frequencies and $r$ is the lattice distance between two oscillators, or spins.
For $\sigma=0$ equilibrium correlations in the low-$T$ regime, $T<T_{KT}$, decay algebraically in space: 
\begin{equation}\label{eq:CrKT}
C_0(r)\sim r^{-\eta(T)}\, , \  \eta(T)=T/2\pi\,,
\end{equation} 
a signature of quasi-long-range order. 
For $\sigma>0$,  it has been shown  that  steady state  correlations decays exponentially in space \cite{RouzaireLevis} 
\begin{equation}\label{eq:Crs}
C_{\sigma}(r)\sim e^{-r/\xi_{\infty}},
\end{equation}
at any finite temperature, signaling the absence of quasi-long-range order in the system.
 
We show in \textbf{Figure~\ref{crt}}\textbf{A}  the spatial decay of $C(r,t)$ at different times following a quench to $T=0.4$, for $\sigma^2=0$ (in purple) and for $\sigma^2=~0.1$ (in green) . The data is shown in a log-log scale to easily discriminate between algebraic and geometric decays. 
As time goes on, spatial correlations build up in the system, indicating the growth of ordered structures, to finally collapse on the before mentioned long-time behaviour. 
To study the coarsening, or phase ordering, dynamics, we define a correlation length $\xi(t)$, extracted from $C(\xi(t),t) = 1/e$ and plot its time evolution in \textbf{Figure~\ref{xi}}\textbf{A} for different values of $\sigma$. 
For $\sigma=0$, we recover without surprise the well-known scaling $\xi(t) \sim \sqrt{t/\log t}$ \cite{rojas1999dynamical, YurkeHuse1993,JelicCugliandolo}. The logarithmic correction takes root in the logarithmic dependence of a defect mobility on its size; note that it is only relevant for large $t$. Such logarithmic correction to the usual $\sim \sqrt{t}$ growth law in systems with non-conserved order parameter dynamics also appears in all related quantities such as the evolution of the number of vortices $n$ or the mean square displacement (MSD) of a single  vortex, as we shall discuss in more detail later on. 
As shown in \textbf{Figure~\ref{crt}}\textbf{B} the dynamic scaling hypothesis \cite{Bray2000} is consistently fulfilled for $\sigma=0$, as all the curves $C(r,t)$ collapse into a single master curve in the scaling regime, once the space variable has been rescaled by the characteristic growing length. 

For $\sigma>0$, the results qualitatively differ: independently of the system size $L$, the growing length $\xi(t)$ saturates at a finite value $\xi_\infty$ at long times. Such steady value decreases as $\sigma$ increases. This is consistent with the limit case $\sigma\to \infty$, corresponding to a system of decoupled oscillators exhibiting coarsening and thus a diverging correlation length at long times, and also with the behaviour of  correlation functions Eq.~(\ref{eq:CrKT}) and (\ref{eq:Crs}). Interestingly, \textbf{Figure~\ref{xi}}\textbf{B} shows that all curves for $\sigma>0$ collapse into a single master curve if both distances and time are rescaled. This master curve follows $f(x) = a(1-\exp(-b\sqrt{x})$, from which we conclude that 
       \begin{equation}
       \label{xieq}
       \xi(t) = \frac{a}{\sigma}\, \left( 1 - e^{-b\sqrt{\sigma^2\,t}}\right) \ ,
       \end{equation}
with $a =  2.3$, a constant related to the long time limit of $\xi$, and $b=0.65$. Note that for small $t$, one can expand $\xi(t)$ to the first order and obtain $\sigma\,\xi(t) \approx ab\,\sqrt{t})$ ; $ab=1.5$ thus represent the slope of the initial $\sim \sqrt{t}$ coarsening.   
From Eq.~(\ref{xieq}) we first conclude that, the limit $t\to\infty$ implies that $\lim_{t\to\infty}\xi(t)=\xi_{\infty} \sim 1/\sigma$, as found in \cite{RouzaireLevis} using a simplified noiseless 1D argument, where it is assumed that the phase difference $\theta_i - \theta_j$ is entirely determined by the frequency difference $\omega_i - \omega_j$ -- a very crude yet informative reasoning. In contrast with the coarsening dynamics of the random field XY model \cite{RFXY}, here we obtain a finite correlation length at long times as soon as our quench disorder is turned on . 
Second, one can extract a scaling relation for typical time $\tau$ associated to the saturation of $\xi(t)$: after a relaxation time $\tau$ following the quench, the growth process stops. 
Indeed, even when the system is driven away from equilibrium, its short-time dynamics follows the passive XY scaling, up to a transient time $\tau$ after which $\xi(t)$ saturates at a finite value $\xi_{\infty}$ . We define $\tau(\sigma)$ as the time for which the equilibrium correlation length is equal to the steady state out-of-equilibrium one:
       \begin{equation}
       \xi(\sigma=0,\tau) = \xi(\sigma,t\to\infty) \ .
       \label{eq:tau}
       \end{equation} 
We plot it in the inset of \textbf{Figure~\ref{xi}}\textbf{B} and obtain $\tau \sim 1/\sigma^2$: the wider the frequency distribution is, the faster is the relaxation process. This intuitive tendency can be easily  captured by the following scaling argument.
As the logarithmic correction is a long term effect, let's drop it for now and assume for the sake of simplicity that $\xi$ first follows an equilibrium growth $\sqrt{t}$. As it eventually saturates at a steady state value $\xi_{\infty} \sim 1/\sigma$, the crossover between both regimes thus occurs when 
\begin{equation} 
       \label{W}
       \sqrt{\tau} \sim \xi_{\infty} \sim 1/\sigma \Rightarrow \tau \sim 1/\sigma^2 \ ,
\end{equation}
confirming the scaling naturally contained in Eq.~(\ref{xieq}). \newline

We complete the picture by reporting the total number of defects $n(t)$ for different driving strengths $\sigma$ in \textbf{Figure~\ref{n}}\textbf{A}.
At equilibrium (purple), one obtains $n(t) \sim \log t\,/\,t$. 
The similarity of the equilibrium long-term scaling for $\xi(t)$ and for $n(t)$ is not a coincidence. The correlation length $\xi$ is intimately related to the total number of defects $n$ in the equilibrium XY model. Indeed, as a first approximation, $\xi$ is given by the average distance between defects; the system being homogeneous, it follows that $\xi \sim 1/\sqrt{n}$. We indeed recover that at equilibrium, $\xi\sqrt{n}$ is constant over time, cf. \textbf{Figure~\ref{n}}\textbf{B}. For $\sigma>0$, $n(t)$ eventually saturates at a finite steady state value, as the number of defects at a given temperature increases with $\sigma$ . This departure from the  XY equilibrium behavior is also clearly reflected by the fact that  $\xi$ is no longer given by the typical distance between defects, as shown in \textbf{Figure~\ref{n}}\textbf{B}. The steady state of the Kuramoto model is characterized by both a finite number of defects and a finite correlation length which are, \emph{a priori}, unrelated. As such, the quench dynamics of the system is no longer fully governed by a single growing length, as its equilibrium counterpart, and the dynamics of the vortices have to be considered in more detail. 

\subsubsection{Topological Defects}
\begin{figure}[]
	 	\centering
		\includegraphics[width=0.85\linewidth]{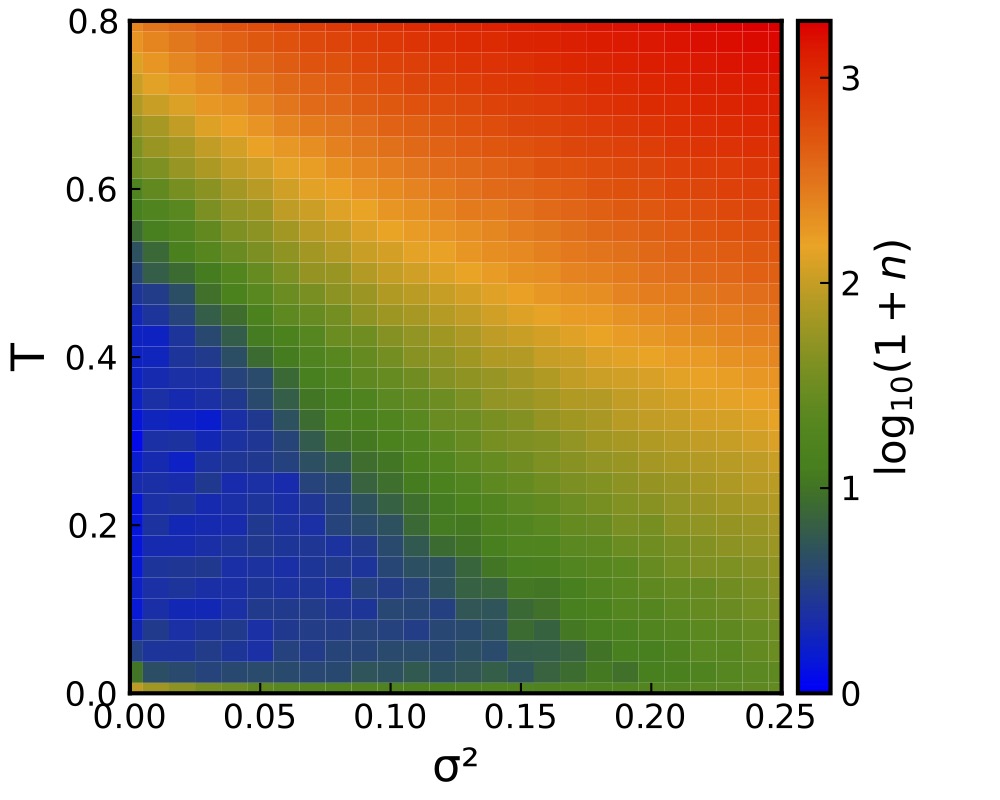}
		\caption{Color map of the number of the number of defects $n$ (in log-scale, $\log_{10}(1+n)$) over the parameter space $\sigma^2-T$ in the steady state. In the blue region  defects are scarce in the system, while in the red one the system is populated by  lots of defects (the number growing exponentially fast as one leaves the blue region). }
		\label{n_phasespace}
	\end{figure}

As a useful reference for our analysis,  we first report a map of the total number of vortices in the steady state over the $\sigma^2-T$ plane in \textbf{Figure~\ref{n_phasespace}}.
As $T$ and $\sigma^2$ decrease, the dynamics gets slower and slower, up to the $T=0$ limit case where the self-spinning alone cannot help annihilating pairs of defects: the system initially disordered remains stuck in a metastable state with a lot more vortices than the number of vortices it would exhibit if relaxed from an initially ordered state. 

\begin{figure}[]
	\includegraphics[width=0.99\linewidth]{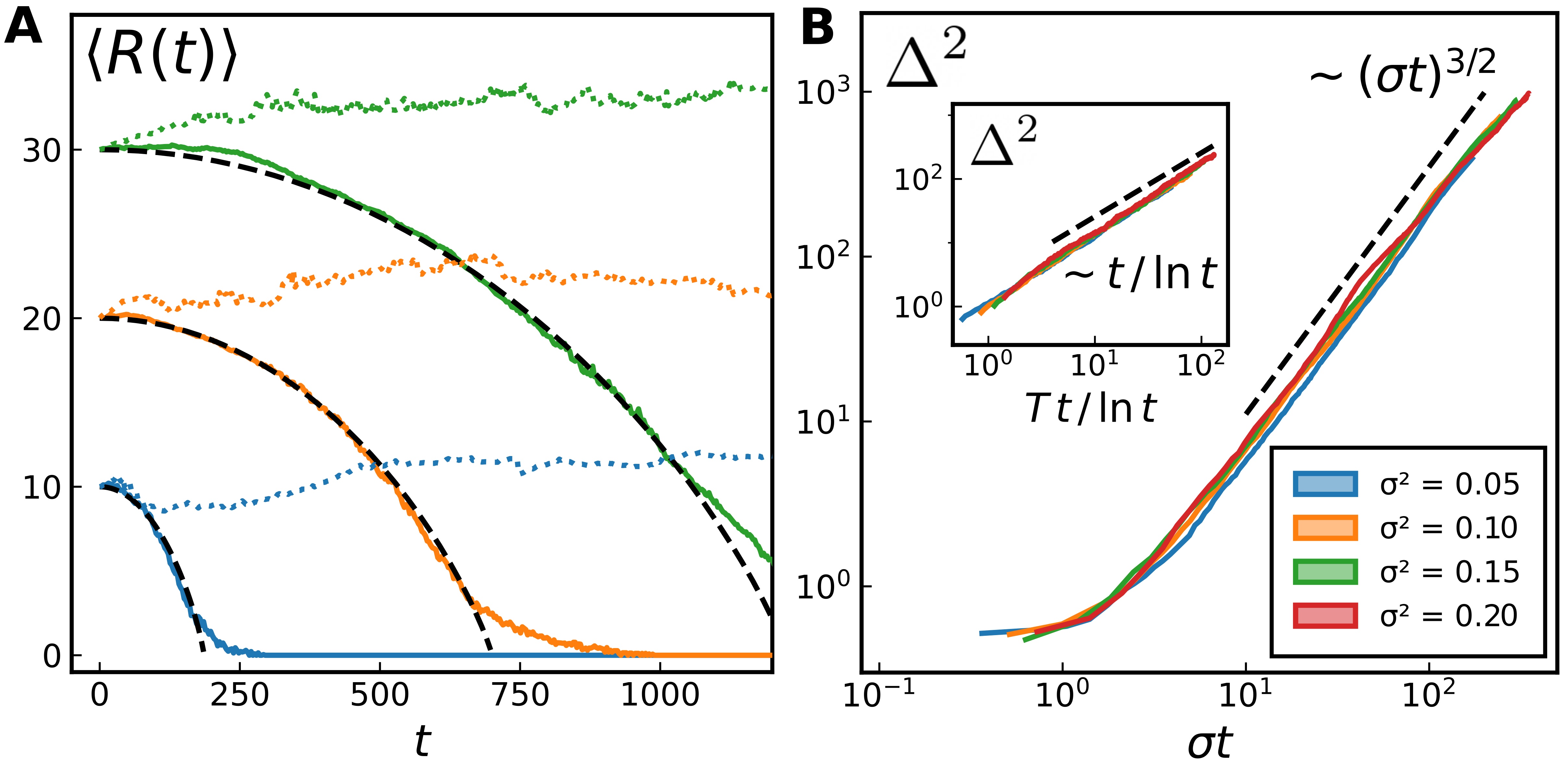}
	\caption{
	\textbf{(A)} Vortex-antivortex separation, averaged over 120 independent runs, for three different initial vortex-antivortex distances  $R_0$ (blue: $R_0=10$, orange: $R_0=20$ and green: $R_0=30$). Full colored lines correspond to $T=0.1$ and $\sigma^2=0$. Dotted colored lines correspond to $T=0.1$ and $\sigma^2=0.1$. Black dashed lines are the predictions for an overdamped dynamics with a Coulomb interaction potential Eq.~(\ref{eq:yurke}). 
	\textbf{(B)} Mean square displacement (MSD) $\Delta^2(t) = \langle (\bold{r}(t) - \bold{r}(0))^2 \rangle$ against time, at $T=0.05$ for different $\sigma$. \textbf{Inset}: MSD for the XY model ($\sigma=0$) for different temperatures ($T=0.2, 0.3, 0.4, 0.5$ from blue [left] to red [right]).}
		\label{MSD}
	\end{figure}

We state once again that the crossover from the defect-rare (blue region) to the defect-rich (red region)  regime does not result from an actual phase transition: there is no qualitative change in the correlation length as one moves across the phase space (the system only exhibits short-range order at any $\sigma\neq0$ for all $T\ge 0$).
At contrast with the standard literature of the XY model \cite{Kosterlitz1973,Cugliandolo}, we do not classify vortices into free and bounded here. Indeed, as mentioned above and discussed in more detail below, the core mechanism responsible for the BKT phase transition, namely the defects' unbinding at \textit{a finite} temperature, breaks down upon self-spinning. 
In the driven model,  topological defects are found to be genuinely free at \textit{any} temperature $T>0$. In other words, there exists no finite temperature at which vortices are bounded into pairs.
While two XY defects of opposite charge attract each other with a logarithmic potential, we find that the defects, upon self-spinning, generically unbind at any $T\ge0$ and $\sigma>0$. We prove it by tracking vortices in time, and considering the average distance between two defects. To do so, we initialize the system in a configuration with a vortex-antivortex pair at a distance $R_0$, enforcing PBC (cf. SM \cite{SM} for details). We then let the system evolve at low-$T$ and low-$\sigma$ such that there is no (or very few) spontaneous creation of vortices. We monitor the distance $R(t)$ between our two defects and present the results in \textbf{Figure~\ref{MSD}}\textbf{A}.

As shown in \textbf{Figure~\ref{MSD}}\textbf{A},  the XY model  data is in perfect agreement with the overdamped description 
\begin{equation}\label{eq:yurke}
\dot{R}\log R = V'(R)
\end{equation}  
where $V(R) \propto \log R$, a Coulomb 2D potential  \cite{YurkeHuse1993}. 
For the Kuramoto model, the vortex-antivortex distance $R(t)$ remains in average equal to their initial distance $R_0$, proving that there is no distinct effective potential between them. 
One could naively attribute this feature to the non-equilibrium nature of the system. However, deciphering the impact of  a non-equilibrium drive on the long-range properties of a system on general grounds is a challenging task and one is usually constrained to rely on specific examples where this question has been addressed. 
For instance, it has been shown that the general BKT scenario remains valid (at least not so far from equilibrium) for an XY model with exponentially correlated thermal noise \cite{Paoluzzi2018} and for the 2D solid-hexatic transition of self-propelled disks \cite{LinoDefects}. In the present model system though, the BKT scenario breaks down. 
	
Another remarkable property of  topological defects upon self-spinning is found in the statistics of their displacement: they feature anomalous diffusion, their mean-square displacement $\Delta^2(t)\sim t^{\alpha}$, with $\alpha=3/2$ (i.e. super-diffusion). Once again we investigate the dynamics of vortices by tracking them. Since, as discussed earlier, defects are free, we can now focus on the motion a single defect. To do so, we prepare an initial configuration hosting a single $+1$ defect at the center of the simulation box. One can exclusively focus on $+1$ vortices without loss of generality because the equation of motion (\ref{eq:model}) is statistically invariant under a change of variable $\theta\to-\theta$ (as long as the distribution of the $\{\omega_i\}$ is centered) and such a transformation transforms a $+1$ vortex into a $-1$ anti-vortex. The dynamics of $\pm 1$ defects are thus statistically identical. We then let the system evolve at low-$T$ and low-$\sigma$ ensuring that (i) new defects are not spontaneously created and (ii) the simulation box is large enough such that boundary effects are negligible. 
The mean square displacement (MSD) thus obtained is shown in \textbf{Figure~\ref{MSD}}\textbf{B}. For the XY model (inset of \textbf{\ref{MSD}}\textbf{B}) and at long times, it follows
	
	\begin{equation}
		\Delta^2(t)=\langle (\bold{r}(t) - \bold{r}(0))^2 \rangle \sim t\,/\,\log t    	 \ ,
	\end{equation}   
	showing the expected correction to the normal diffusion scaling, due to the logarithmic dependence of the defect mobility on its size \cite{YurkeHuse1993} [$\bold{r}(t)$ being the position of the vortex in the lattice at time $t$.].
	
	Upon self-spinning, the defects become super-diffusive and the MSD (cf. \textbf{Figure~\ref{MSD}}\textbf{B}) instead follows
	\begin{equation}
		\Delta^2(t)  \sim (\sigma t)^{3/2} \ .
	\end{equation}  
	The $\sigma$-dependence of the vortex dynamics enters through a rescaling of time: the typical time associated with its motion along a domain is proportional to its typical size $\xi \sim 1/\sigma$. A careful characterisation of the stochastic motion of the defects will be presented in \textbf{Section~\ref{sec:defectsdynamics}}.
	

	\subsection{Underdamped Dynamics} \label{sec:underdamped}
The nature of the dynamics, being overdamped or underdamped, does not have any impact on the long-time, large-scale properties of a system as long as its dynamics fulfill detailed balance. If this defining feature of equilibrium is satisfied, the steady-states obey  Boltzmann statistics. For the Kuramoto model, being out-of-equilibrium, different ways of exploring the configuration space might lead to different steady large scale behaviour \cite{gupta_kuramoto_2014,komarov_synchronization_2014,olmi_hysteretic_2014}. We are thus interested in putting into test the robustness of our results to the presence of inertial effects.

\begin{figure}[]
		\includegraphics[width=0.99\linewidth]{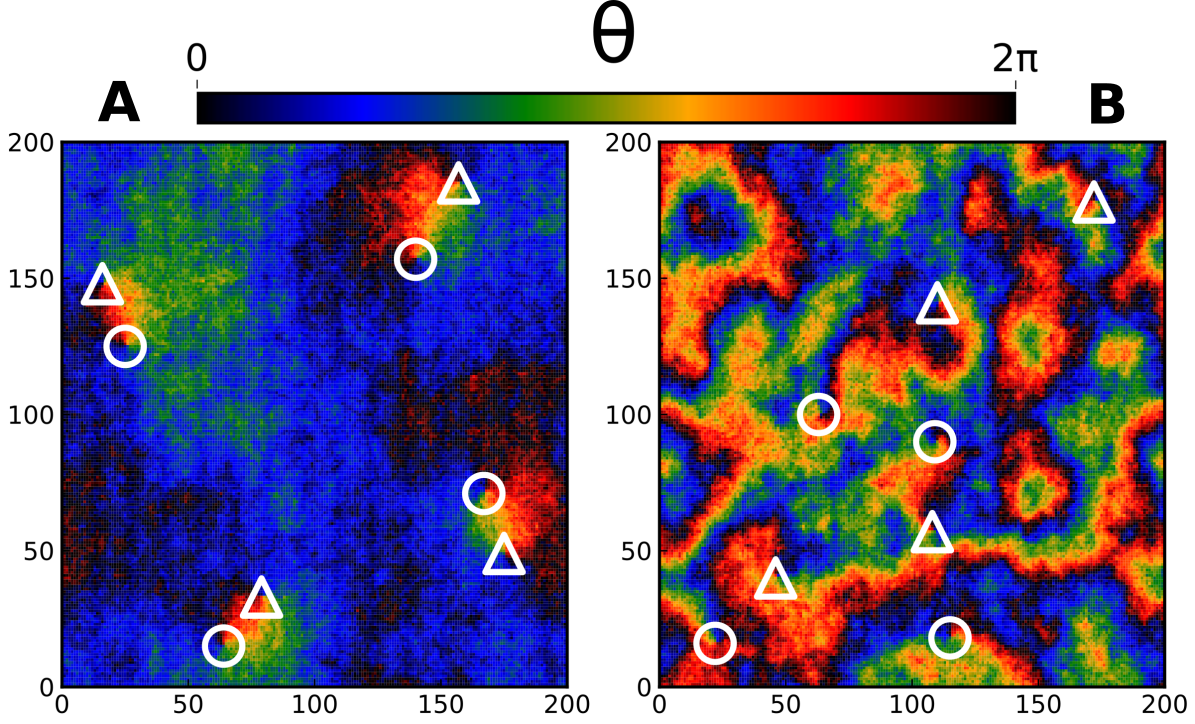}
		\caption{
			Snapshots for $m=1$, $T =0.2$, \textbf{(A)} $\sigma^2=0$ and \textbf{(B)} $\sigma^2=0.1$. The phase of each oscillator is represented by a color scale. Vortices (antivortices) are represented by circles (triangles).
		}\label{snapshots_m}
	\end{figure}

We first explore the steady states produced by letting the model Eq.~(\ref{eq:under}), with inertia,  relax from an initially disordered state with many vortices, to a state with only a few of them  (i.e. the low temperature regime). If again, we look at typical  snapshots of the steady state, \textbf{Figure~\ref{snapshots_m}} (for the same parameter values in \textbf{Figure~\ref{snapshots}} but $m=1$), we observe basically the same features as in the overdamped case: for $\sigma=0$ we find oppositely charged defects pairs visually connected by a phase field reminiscent of the magnetic field lines generated by a planar magnet, while for  $\sigma>0$ vortices are surrounded by a phase texture that makes difficult pairing them by eye, featuring synchronised domains of different shapes and sizes.

	\subsubsection{Coarsening Dynamics}
	
\begin{figure}[]
		\includegraphics[width=0.99\linewidth]{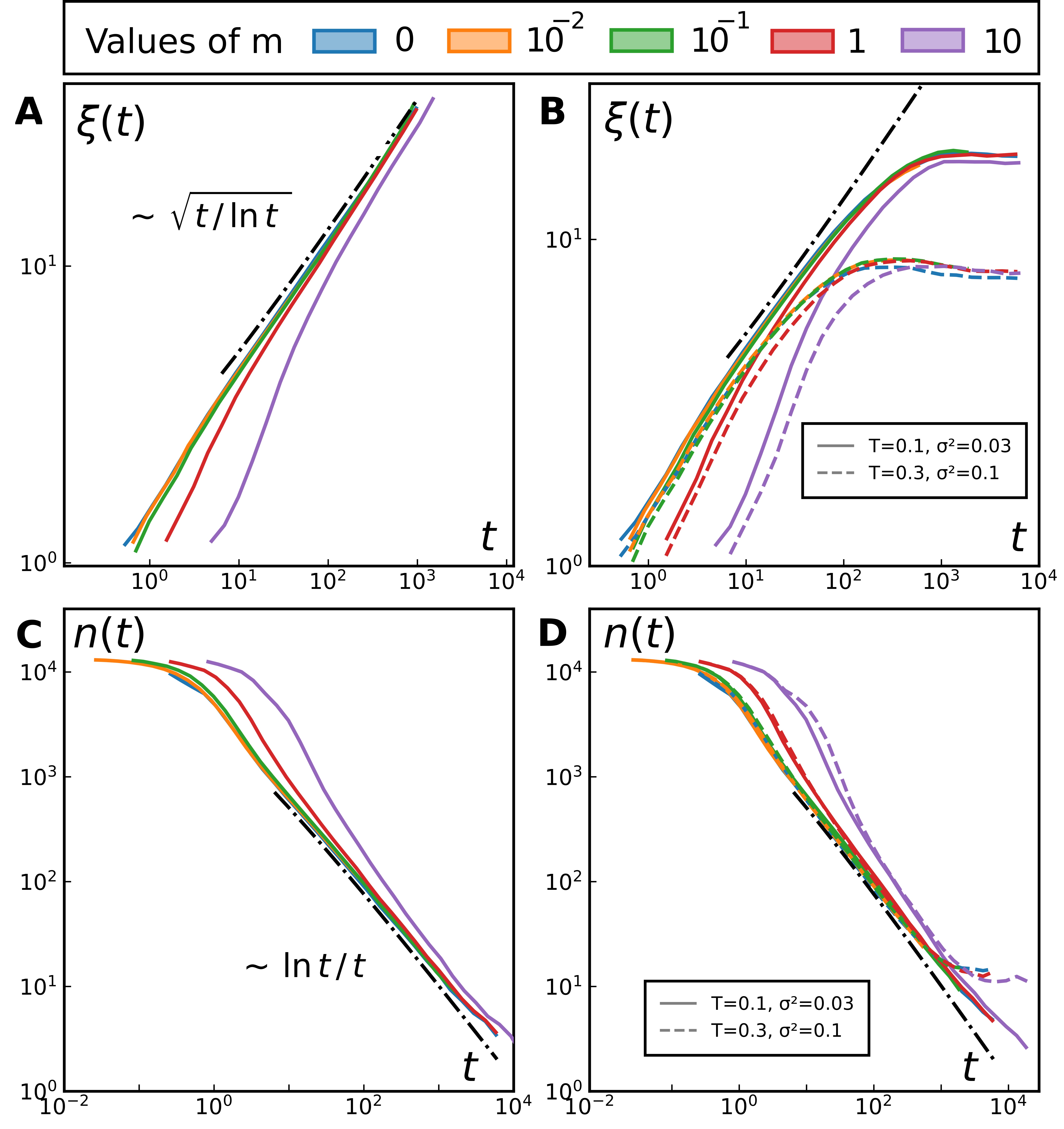}
		\caption{Relaxation from an initially disordered state. Different colours represent different inertia $m$ for all four panels, as indicated at the top of the figure. \textbf{Top row:} Time evolution of the correlation length $\xi(t)$ for \textbf{(A)} the XY model at $T=0.2$ and  for \textbf{(B)} the driven system at the different values of $T$ and $\sigma$. \textbf{Bottom row:} Time evolution of the total number of defects $n(t)$ for \textbf{(C)} the XY model at $T=0.2$ and for \textbf{(D)} the driven system at the different values of $T$ and $\sigma$. We only display two sets of parameters for readability but we have reached the same conclusions for all the other tested parameters (distributed across the entire phase space). \textbf{Left column:} At equilibrium. \textbf{Right column:} out-of-equilibrium.}
		\label{xinm}
	\end{figure}

We now turn into the relaxation dynamics of the model with inertia following a quench from an initially disordered state.	
We investigate the same dynamical quantities as for the overdamped regime (cf. \textbf{Figures~\ref{crt},\ref{xi}} and \textbf{\ref{n}}), varying the inertia from $m=10^{-2}$ to 10 and comparing them to their overdamped counterparts (in blue, $m=0$). The results are reported in \textbf{Figure~\ref{xinm}}. 

	At equilibrium, the results confirm that we do recover the same long term values for $\xi$ and $n$ independently from the inertia $m$, cf. \textbf{Figure~\ref{xinm}A,C}. The results also indicate that, even in the presence of self-spinning, the inertia does not affect significantly the  long-time behavior as one recover the same steady state values for values of $m$ covering three orders of magnitude (see \textbf{Figure~\ref{xinm}B,D}). The short-time dynamics is however affected by the inertia of individual rotors.

	\subsubsection{Topological Defects}
\begin{figure}[]
		\includegraphics[width=0.99\linewidth]{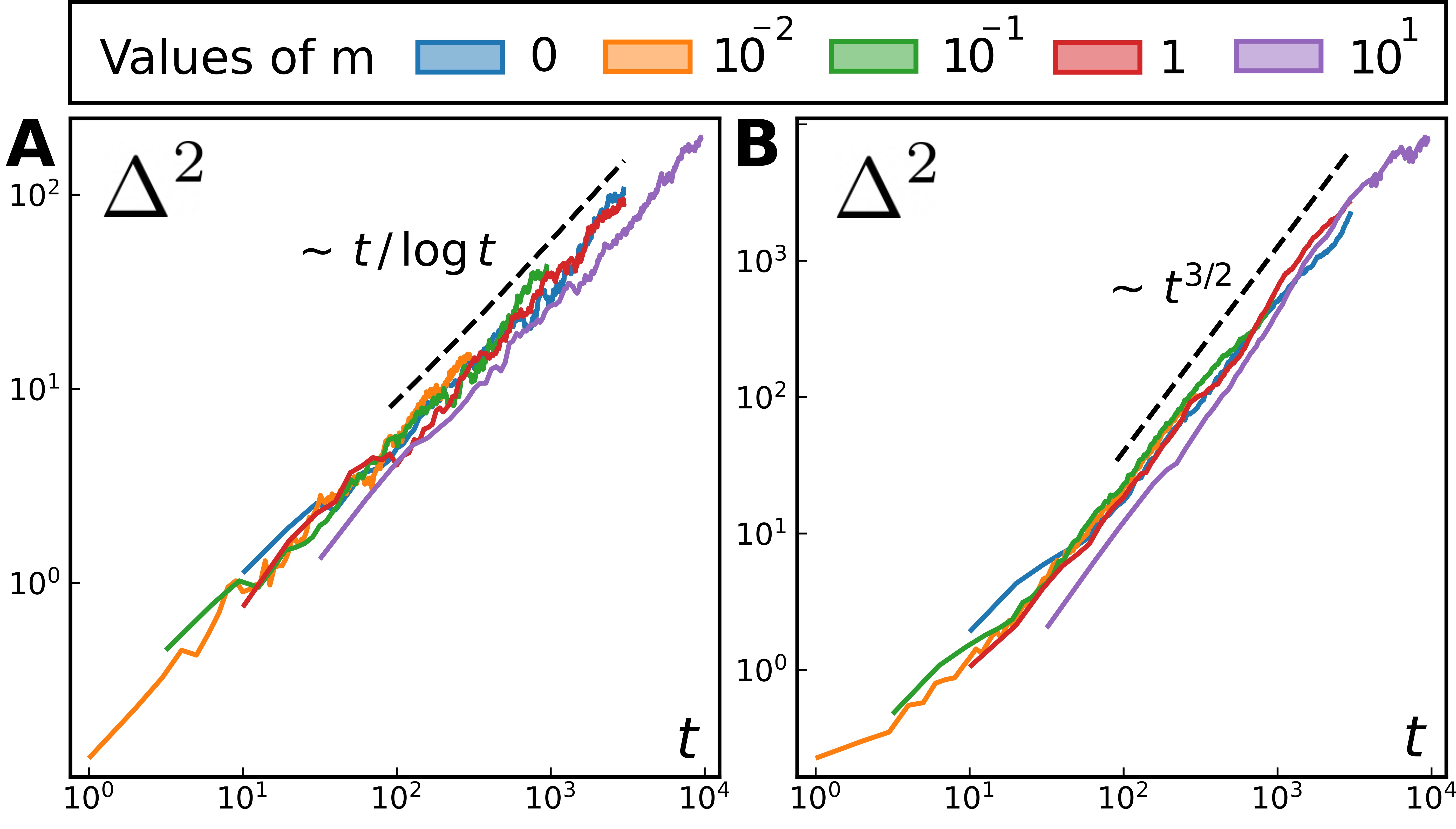}
		\caption{Mean Square Displacement (MSD) of individual defects for \textbf{(A)} XY model at $T=0.2$ \textbf{(B)} forced model at $T=0.2$ and $\sigma^2=0.03$ .}
		\label{MSD_m}
	\end{figure}

	We now turn back to topological defects and shall see that, surprisingly, their MSD is not affected by the inertia of individual oscillators. We resort to the same protocol (track a single defect over time in a big system with free boundaries) and plot the result in \textbf{Figure~\ref{MSD_m}}. 
	We recover the same (sub)diffusive motion for the XY model and the super-diffusion with the same exponent in the presence of self-spinning. 
	
	Overall, our results are robust as they indicate that the long-term properties of the system are not affected by the inertia of the underlying individual oscillators.
	We emphasis that this result is new and somehow surprising as, beyond the fact that there are no general guidelines for out-of-equilibrium systems, including inertia in a model of coupled oscillators can yield dramatic changes. For instance, in the \textit{globally} coupled Kuramoto model inertia changes the order of the disordered/synchronised transition from second to first order \cite{tanaka_first_1997,gupta_kuramoto_2014,olmi_hysteretic_2014,komarov_synchronization_2014}. It is therefore remarkable that this does not happen in a model of  \textit{locally} coupled oscillators.

	And finally, we have seen that a non-equilibrium forcing $\sigma > 0$ dramatically changes the MSD regime from sub-diffusive to super-diffusive.

	\section{Discussion} \label{sec:discussion}
	\subsection{The fate of the BKT transition}\label{sec:collapse}
	

	The BKT transition scenario usually occurs when topological defects interact with a $\sim 1/r$ force, in 2D, establishing quasi-long range order at low temperatures. After a quench across such a phase transition, due to the collective behaviour of topological defects, the typical length scale grows as $\xi \sim \sqrt{t/\log t}$. Such behaviour is driven by the  diffusion and annihilation of defects, whose number decreases as $n \sim 1/\xi^2$. 
	
	In the presence of intrinsic spinning, the field disturbances generated by the defects only expand over relatively short length scales, at any temperature. This can be understood from the equation of motion defining the model. 
	Consider Eq.~(\ref{eq:model}) at $T=0$ for a test spin at distance $r$ from a topological defect. In the absence of self-spinning, the force  generated by the defect is $\sum \sin (\Delta \theta) \sim 1/r$. Incorporating such term in the equation of motion, one gets $\dot{\theta_i} = \sigma \omega_i + 1/r$. From this simplified expression, it is easier to understand that the influence of the defect is dominant \textit{only} within a radius $\xi \sim |\sigma \omega_i|^{-1}$ which, averaged over all spins, boils down to $\xi \sim 1/\sigma$ as $\omega_i$ has unit variance.

	Indeed, the behaviour of a pair of vortices shows a clear absence of any relevant defect-defect interacting potential: topological defects are genuinely free in the driven system. Once the phase patterns have been formed, they screen the expected $\sim\log \ r$ potential of the XY model, key prerequisite to belong to the BKT universality class. 
	The measurements of the correlation length $\xi \sim 1/\sigma$ and the absence of relation between $\xi$ and $n$ upon self-spinning strengthen that claim. It only took an energy injection at the smallest scale, independent from spin to spin, for the BKT scenario of the XY model to collapse. As it shows that long-range influence is lost as soon as the model is made active, the present work conceptually supports the conclusions of Pearce \textit{et al.} \cite{pearce_orientational_2021}, where they report the absence of long-range ordering of defects in active nematics.

	However, this out-of-equilibrium system still inherits some of the coarsening dynamics of the 2D XY model physics: \textbf{Figures~\ref{xi}} and \textbf{\ref{xinm}B,D} show that the short-time dynamics do follow  $\xi \sim \sqrt{t/\,\log t}$, $n(t) \sim \log t /t$ and $\Delta ^2 \sim t/\log t$. As the features specific of the non-equilibrium system (the creation of the domain boundaries, for instance) take some transient  time to establish, the underlying equilibrium XY mechanisms prevail at short times. For example, consider an initially disordered system, at first crowded with defects: the vortex mean free time is so small that even though they are intrinsically free, they rapidly end up colliding and annihilating, as bounded vortices would have done. We have shown that such transient time from equilibrium to active dynamics scales as $1/\sigma^2$.  

	\subsection{Defects' dynamics} \label{sec:defectsdynamics}
	
	This section aims at shedding some light on the underlying mechanisms responsible for the spontaneous creation of  domain boundaries and  topological defects, and  their super-diffusive random motion.

	\subsubsection{Creation of domain boundaries and defects} \label{subsec1}
	Interested by the emergence of the patterns such as those in \textbf{Figures~\ref{snapshots}B} and \textbf{\ref{snapshots_m}B}, we look at the short time dynamics of the fields $\theta\equiv\theta(x,y)$ and $\dot{\theta}\equiv\dot{\theta}(x,y)$ ($x$, $y$ being the spatial coordinates of a point in the lattice). We do so by following the evolution of the system from an ordered initial condition (all sites with identical phase). We proceed in such a way in order to isolate the underlying mechanisms at stake and avoid being blurred by the numerous defects inherent to a disordered initial condition. As the time series of $\theta$ is noisy for $T>0$, one has to average in time to obtain a meaningful signal $\dot{\theta}$. Details about the exponential moving average used in that preliminary signal processing step can be found in the SM \cite{SM}.

	\begin{figure}[]
		\centering
		\includegraphics[width=0.99\linewidth]{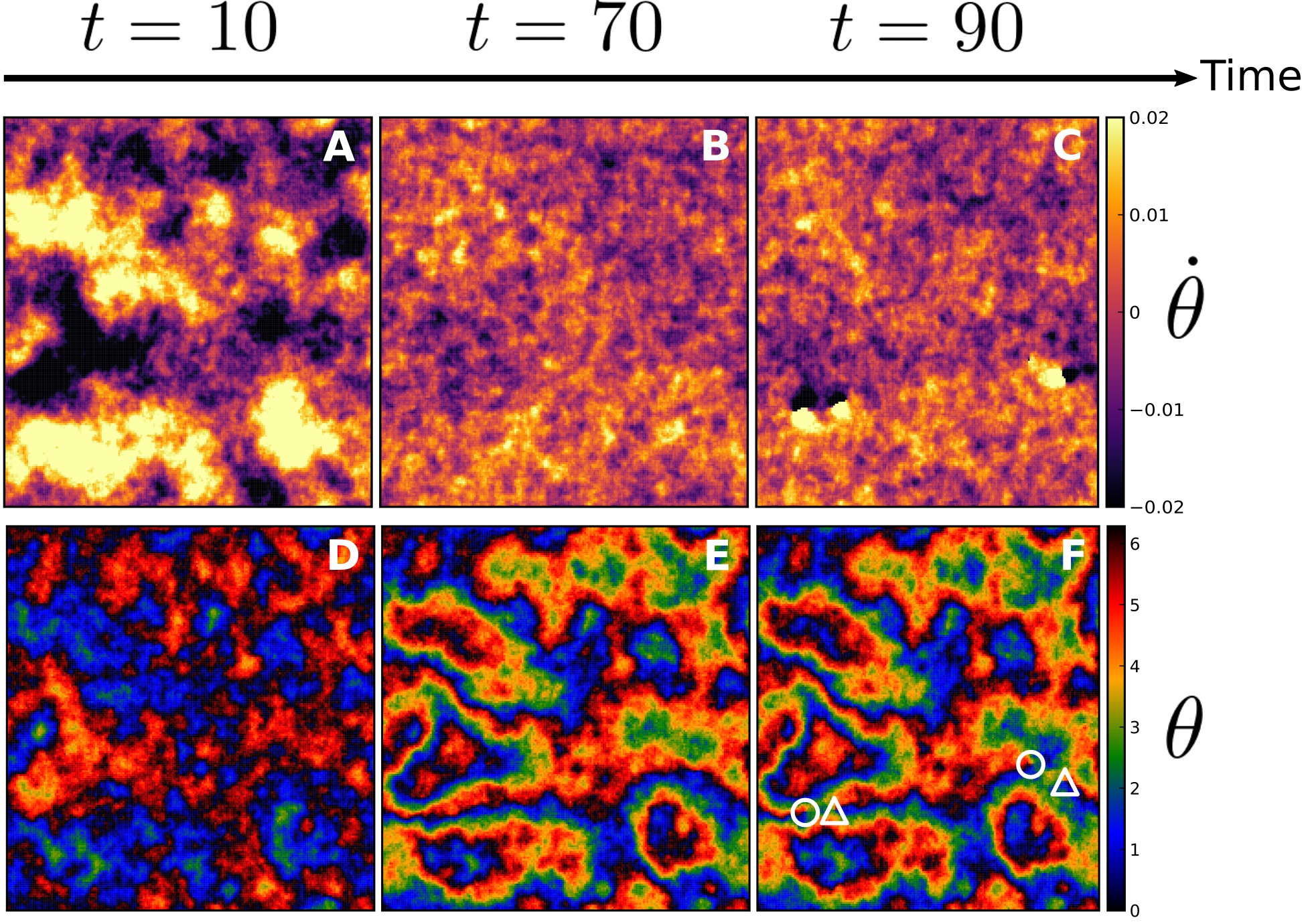}
		\caption{
			Three snapshots over time of the fields $\dot{\theta}(x,y)$ \textbf{(top row)}  and $\theta(x,y)$ \textbf{(bottom row)} for a system initially ordered with $T=0.05$ and $\sigma^2 = 0.2$. In panel \textbf{(C)}, defects are clearly identified thanks to the localized and intense amplitude of the instantaneous frequencies. In panel \textbf{(F)}, $+1$ defects are highlighted by a circle, $-1$ defects are highlighted by a triangle. 
		}\label{snapthetadot}
	\end{figure}

	After very few time-steps, a clear picture appears: locally synchronised regions develop in the $\dot{\theta}$ field. We recall that since $d=2$ is the lower critical dimension for frequency locking in the noiseless short range Kuramoto model, we do not expect more than local frequency synchronization. Since the average instantaneous frequency (over the $N$ spins at a given time $t$) scales as $1/\sqrt{N}$ ---as one should expect from a zero-mean distribution of intrinsic frequencies, reciprocal interactions and gaussian white noise---, some regions rotate clockwise and some others counterclockwise; look for instance at the dark and bright regions of \textbf{Figure~\ref{snapthetadot}}\textbf{A}. Naturally, these locally synchronized regions in $\dot{\theta}$ translate  into locally aligned domains in the phase field. This is illustrated in \textbf{Figure~\ref{snapthetadot}}:  the dark region by  the left of panel \textbf{\ref{snapthetadot}A} generates a phase synchronised domain of same shape and at the same position in panel \textbf{\ref{snapthetadot}D}, now in red.

	Progressively, as time goes on,  the different domains grow, (cf. panel \textbf{\ref{snapthetadot}E}), defining a network of  domain boundaries. As they correspond to regions where $|\nabla \theta|$ is large, they can be visually identified by looking for thin elongated regions across which colours change rapidly (though in a smooth way in contrast to, for instance, domain boundaries in the Ising model).
	At this stage, the domains are locally ordered, their boundaries concentrate most of the energy of the system and  the instantaneous frequencies $\dot{\theta}$ gradually decrease in amplitude (compare panels \textbf{\ref{snapthetadot}A} and \textbf{\ref{snapthetadot}B}). The system finally reaches its steady state characterised by synchronised domains resulting from  the competition between the elastic energy and the driving frequencies $\sigma \omega_i$.

	These domain boundaries play a crucial role in the formation of defects. There, the important gradients $|\nabla \theta|$ provide most of the energy needed to create and sustain topological defects. The remaining energy contribution eventually comes from thermal fluctuations, explaining why the temperature threshold necessary to spontaneously generate defects decreases as the forcing increases, see \textbf{Figure~\ref{n_phasespace}}. It also explains why one observes defects creation primarily at the domain boundaries, as exemplified in \textbf{Figure~\ref{snapthetadot}}\textbf{F}. 
	This creation mechanism contrasts with that of the equilibrium case, entirely due to thermal fluctuations and hence homogeneously distributed in space. 
	Yet, the creation mechanism is not the only difference between the active and the passive case: a major difference lies in the motion of these topological defects, as we detail in the last part of this article.


\subsubsection{Defects Super-diffusion} \label{subsec3}

\begin{figure*}[]
	\centering
	\includegraphics[width=0.9\linewidth]{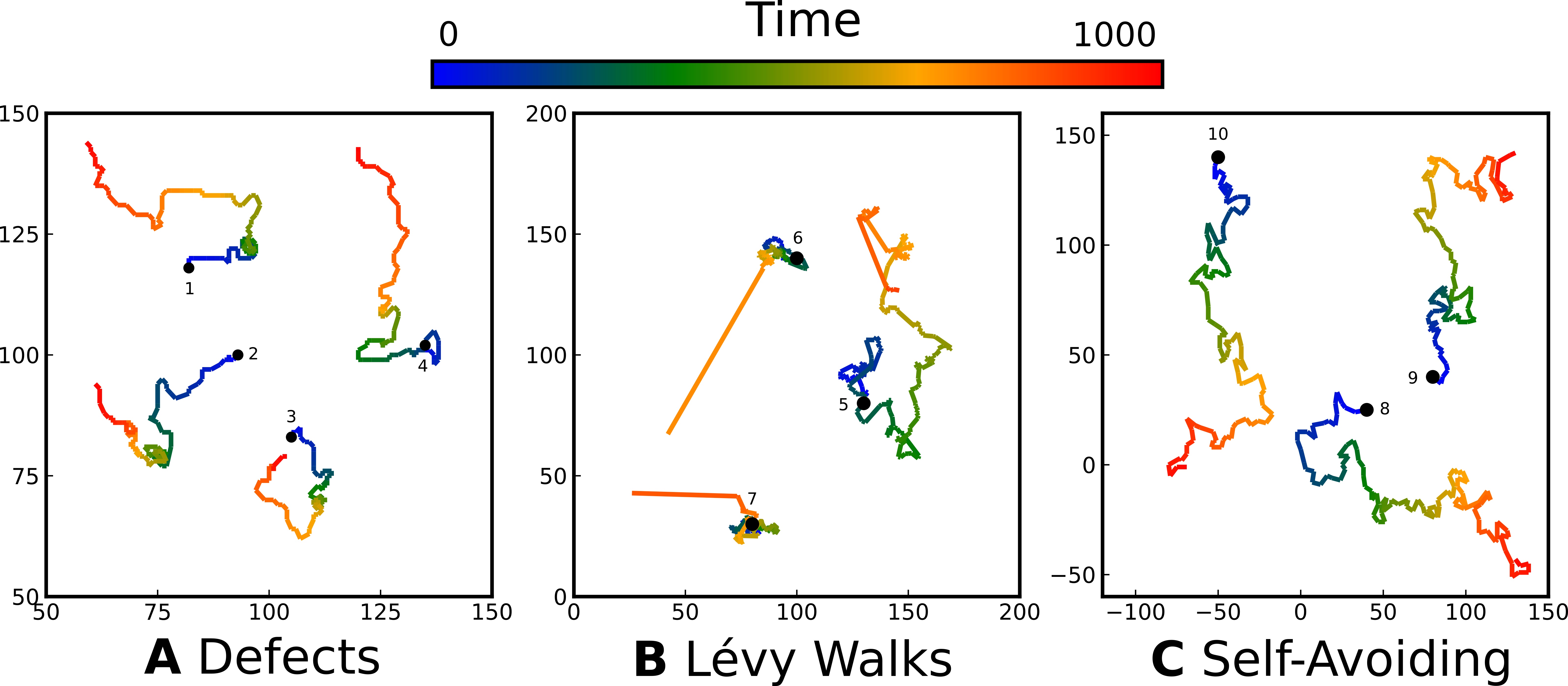}
	\caption{Sample trajectories of \textbf{(A)} topological defects in the actual spin model \textbf{(B)} L\'evy walks with exponent $\gamma= 3/2$ \textbf{(C)} self-avoiding random walks (SAW). 
		Each path departs from a black circle and lasts for $\Delta t= 10^3$, indicated by the color code.}
	\label{traj}
\end{figure*}

The MSD of individual defects is shown in \textbf{Figures~\ref{MSD}B} and \textbf{\ref{MSD_m}}. As soon as $\sigma\neq 0$, in the time window allowed by our simulations  (spanning over 4 decades) the MSD of the defects has an anomalous exponent  $\alpha=3/2$ without any sign of a crossover to a diffusive regime at longer times. We depict in \textbf{Figure~\ref{traj}}\textbf{A} typical vortex trajectories. Since super-diffusion is a frequent phenomenon in active matter (e.g. topological defects in active nematics \cite{SaguesRev,Giomi2014,shankar2022topological,Gompper2020} or cells in biophysics experiments \cite{huda_levy-like_2018}), it naturally led to a plethora of random walk models exhibiting transient \cite{caprini2022dynamics,villa-torrealba_run-and-tumble_2020} or long term super-diffusion \cite{fedotov_emergence_2017,han_self-reinforcing_2021,bertoin1996levy}. 

Among models featuring long-term super-diffusion, the family of L\'evy processes is a popular choice, in particular in biology \cite{huda_levy-like_2018,ariel2017chaotic}. In L\'evy walks (respectively L\'evy flights, their discontinuous counterpart), the duration $\tau$ of each walk (respectively the size of each jump) is usually sampled from a fat-tailed distribution $f(\tau) \sim \tau^{-(1+\gamma)}$, where $1<\gamma<2$ is a common choice for super-diffusing L\'evy walks. The resulting MSD follows $\sim t^{3-\gamma}$ (see \cite{levywalks_review} and references therein for a complete review). 
The super-diffusion stems from the infinite variance of this distribution, explaining why the resulting trajectories sometimes feature a dramatic jump (as trajectories 6 and 7 depict in \textbf{Figure~\ref{traj}}\textbf{B}. As such jumps cannot occur in our system, L\'evy processes do not provide a faithful description of the random motion of vortices in the Kuramoto model. In addition, there is no physical argument to support the specific choice of the exponent $\gamma = 3/2$ (the only one reproducing the anomalous exponent of the vortices).

\begin{figure}[]
	\includegraphics[width=0.99\linewidth]{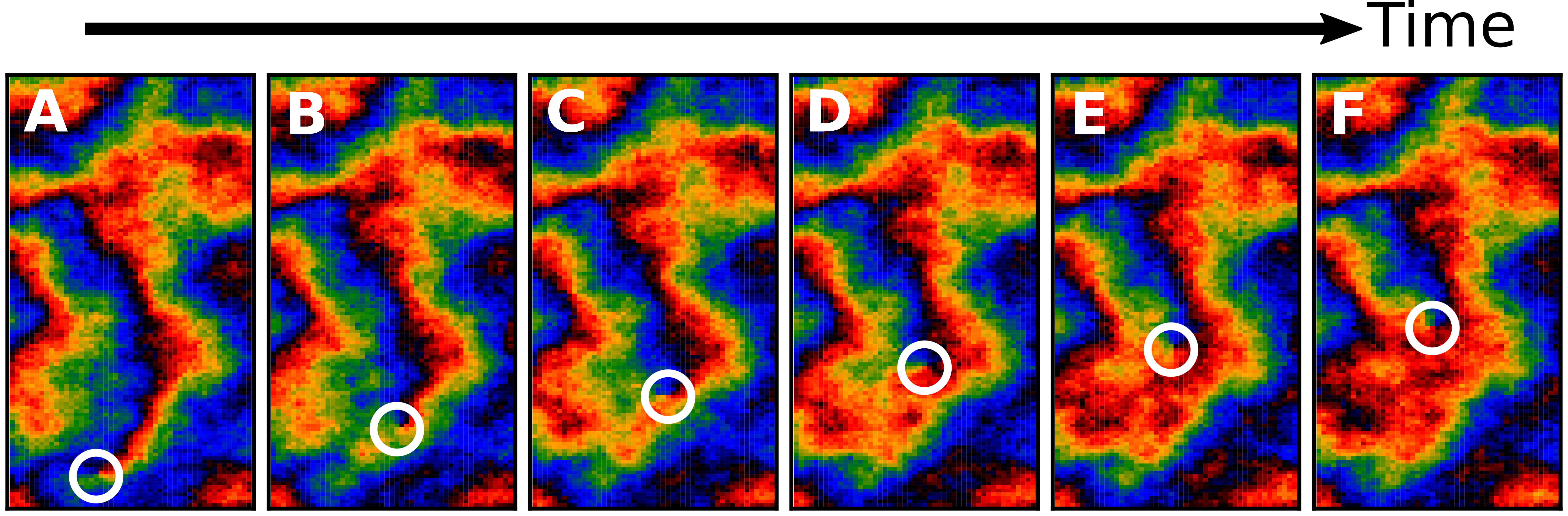}
	\caption{
		Successive snapshots of a fixed 50$\times$120 portion of a 200$\times$200 system evolving with $T = 0.1$ and $\sigma^2 = 0.15$. The topological defect of charge $+1$ is highlighted with a white circle. It moves along the domain boundary (thin black line) and erases it in its wake, leaving an aligned region behind.  
	}\label{surf}
\end{figure}

\begin{figure}[]
	\includegraphics[width=0.99\linewidth]{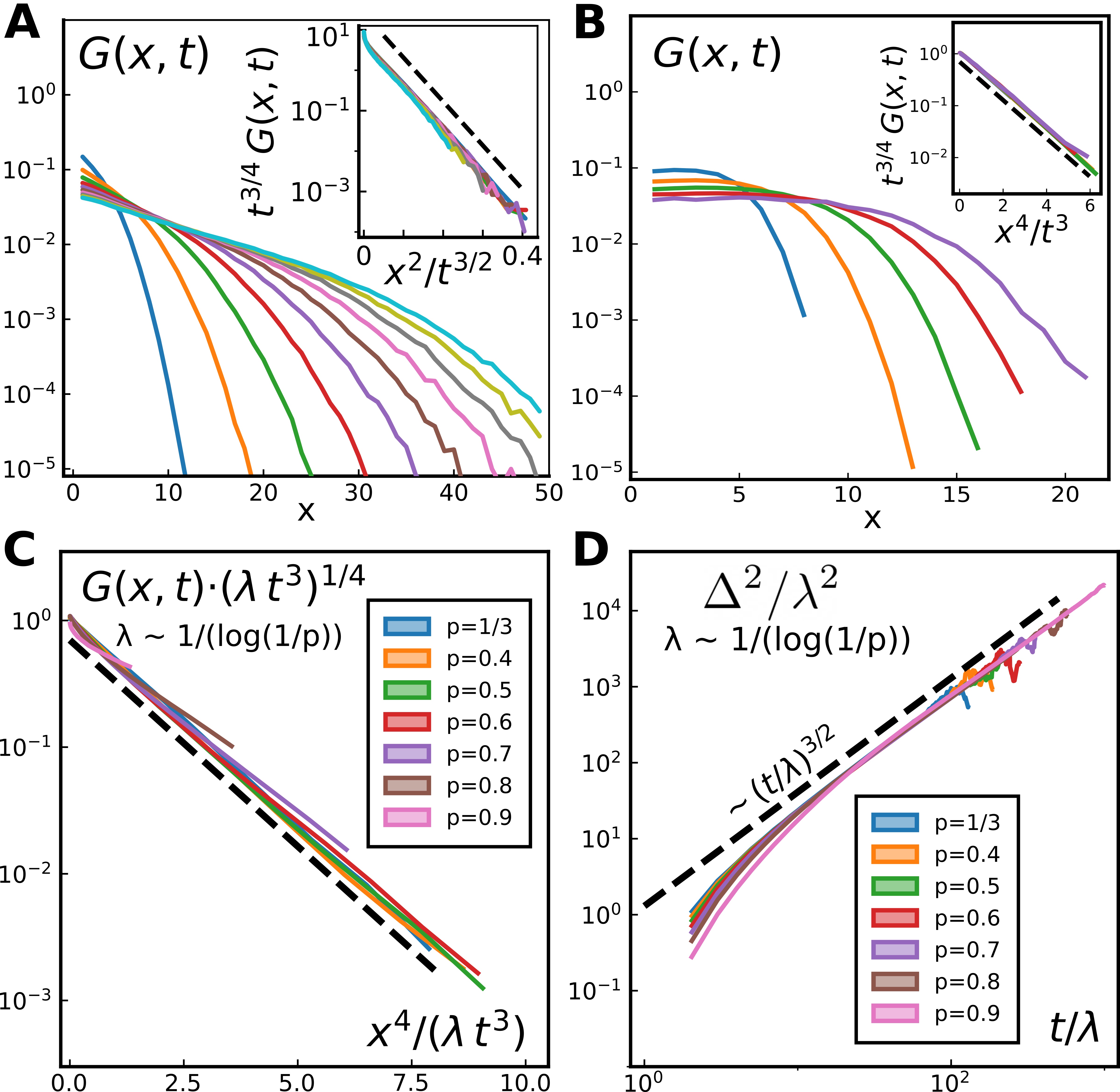}
	\caption{Probability densities of displacement $G(x,t)$ (see main text for definition) for different cases. Both insets represent rescaled data in order to highlight the functional form of $G$. We considered 
		\textbf{(A)} topological defects with $\sigma^2 = 0.025$ and $T = 0.2$, at times (from left to right) $t = 50,100,150,...,500$. Inset: the dashed line follows $f(x) \sim \exp(-60x)$ 
		\textbf{(B)} SAW ($\Delta t = 1$) at times (from left to right) $t = 10,15,20,25,30$ . Inset: the dashed line follows $f(x) \sim \exp(-0.85x)$ 
		\textbf{(C)} Persistent SAW (see main text for definition) with different persistence probabilities on rescaled axes to highlight that the functional form is identical for all $1/3\le p<1$. The dashed line follows $f(x) \sim \exp(-3x/4)$.  
		\textbf{(D)} MSD resulting from $G(x,t)$ of panel (C), for different persistence probabilities $p$. 
	}\label{4panels}
\end{figure}

An important feature of the system to grasp the dynamics of the vortices is the domain structure of the system. 
Indeed, there is a feedback between the spatio-temporal patterns dynamics and the defects' motion. 
On one hand, defects have a strong preference to move along domain boundaries, where elastic energy is the largest  \cite{RouzaireLevis,ChardacBartolo2021}. 
As a domain boundary separates two neighbouring synchronised regions, it is characterised by an excess elastic energy. They thus provide a preferential direction for the motion of the topological defects, in contrast with the isotropic random motion of a single defect in the equilibrium XY model, for instance. 
On the other hand, such excess energy is released during the motion of the defects along  domain boundaries, which are erased as vortices move along them. As shown in the successive snapshots of \textbf{Figure~\ref{surf}}, defects act like a zip, bringing together the two domains on each side of the boundary (e.g. in \textbf{Figure~\ref{surf}A}, these are the two red regions, separated by the thin black line), leaving a synchronized region behind (big red region in \textbf{Figure~\ref{surf}F}). This area is no longer favorable to the future passage of a defect (be it itself or another one), as the elastic energy it previously contained has been dissipated. As such, the phase pattern has a strong impact on the motion of the defects and conversely, defects significantly alter the structure of the system as they move.

As vortices remove the domain boundaries as they move, they perform a random walk with memory: it is more likely for a vortex to keep moving along the domain boundary than retrace its steps. 
The classical framework to treat this kind of motion is the self-avoiding random walk (SAW), which  might be at the origin of the anomalous exponent $3/2$. 
In order to explore whether SAW could provide a useful description of our defects motion, we simulate a few of SAW and visually compare the generated trajectories with the ones we recorded for vortices. As shown in \textbf{Figure~\ref{traj}}\textbf{C}, they look qualitatively similar. 
To be more quantitative, we also computed the MSD and the distribution of displacements $G(x,t)$, defined as the probability that a walker initially at $x=0$ at $t=0$ is at position $x$ at time $t>0$. SAW feature a super-diffusion with $\Delta^2(t)\sim t^{\nu}$ with $\nu = \frac{6}{d+2}$ \cite{barat1995statistics}, hence providing an explanation of the $3/2$ exponent we found for the defects' MSD (here $d=2$, the number of spatial dimensions). Yet, the description provided by SAW is not complete, as  the full distribution of displacements differ significantly.   As shown in  \textbf{Figure~\ref{4panels}}\textbf{A}, for defects, 
\begin{equation}
G(x,t) \sim \exp(-x^2/t^{3/2})
\end{equation} 
while for SAW, one obtains (cf. inset of \textbf{Figure~\ref{4panels}}\textbf{B}) 
\begin{equation}
G(x,t) \sim \exp(-x^4/t^{3})\,.
\end{equation}

We also investigated the effect of including self-propulsion to mimic the behaviour of vortices on domain boundaries. To do so, we add persistence to the walk, on top of the self-avoiding condition. Instead of continuing straight, turning left or turning right with equal probability 1/3, we set the probability to continue straight to $1/3\le p<1$, and thus turning left or right with probability $(1-p)/2$. This process introduces a typical length $\lambda \sim 1/\log(1/p)$ , as $p^n = e^{-n/\lambda}$. The typical timescale before the trajectory has changed orientation with probability 1/3 also scales as $\lambda$. In other words, a persistent SAW, when expressed in terms of the rescaled space $x/\lambda$ and rescaled time $t/\lambda$, is indistinguishable from a non persistent SAW expressed in terms of $x$ and $t$. This explains why, for all $1/3\le p<1$, the probability density of displacement for a persistent SAW with probability $p$ is given by 
\begin{equation}
	G(x,t) = \frac{1}{2\,\Gamma(5/4)\,(\lambda\,t^3)^{1/4}}\ \exp(-\frac{x^4}{\lambda\,t^3})\ , 
\end{equation}
where $\Gamma(x)$ is the usual Gamma function.
Note that the slope of the curves in \textbf{Figure~\ref{4panels}C} determines the exact value of the rescaling factor: $\lambda = 4/(3\log(1/p))$. \newline

Overall, a complete, faithful microscopic description of the topological defects' motion in the  short-range Kuramoto model remains to be found. Despite the simplicity of their displacements' functional form $G(x,t) \sim \exp(-x^2/t^{3/2})$, it is likely that collective effects drive the defects' motion in a way simple random walk models cannot capture, calling for a full many-body treatment of the problem. 

\section{Conclusion}
In this article we have studied the dynamics of the short-range noisy Kuramoto model, constructed as a non-equilibrium extension of the 2D XY model, where spins rotate with an intrinsic frequency, taken from a (quenched) Gaussian distribution.  Exploiting the connection between the emergence of synchronisation and the topological Berezenskii-Kosterlitz-Thouless phase transition in the 2D XY model, we investigate the dynamics of vortices in detail, contributing to our current understanding of the dynamics of topological defects in non-equilibrium soft condensed matter. 

In 2D, the short range noisy Kuramoto does not exhibit any kind of phase transition at any finite temperature, its correlation length is always finite and scales as $\xi\sim\sigma^{-1}$. Instead, there is a crossover between a high temperature region with many topological defects, and a low temperature region with a few ones, as the number of vortices decreases exponentially fast close to such crossover temperature. 
We have showed that the relaxation towards the low temperature regime from a disordered initial state proceeds via the growth of a characteristic length scale, setting the typical size of synchronised domains Such growth does now proceed indefinitely, as the correlation length of the system is finite. The number of defects decays in time following the expected behaviour from the 2D XY model, although the mean separation between defects is  not  given by growing length extracted from the correlation function, meaning that the dynamics cannot be fully described by a single length scale, as for the coarsening of the 2D XY model. 
Indeed, vortices are free in the non-equilibrium model, advected by the domain boundaries of phase synchronised regions, resulting in super-diffusion with a long-time mean-square displacement $\Delta^2(t)\sim t^{3/2}$. Thus, the mean distance between topological defects does not bare any information regarding the large-scale structure of the system.

We have found that our results remain valid both for inertial and over-damped Kuramoto dynamics, showing the robustness of the before-mentioned scenario and the breakdown of BKT physics. We have provided a detailed characterisation of the random walks performed by vortices, and have found that, although their stochastic dynamics shares several similarities to self-avoiding walks, in particular the scaling $\Delta^2(t)\sim t^{3/2}$, their full statistics of displacements are not equivalent. A faithful description of the dynamics of  the dynamics of vortices would require a theory  capturing the two-way dynamical feedback between the phase field and the vortices, a challenging endeavour that we leave for future work.

\section*{Conflict of Interest Statement}

The authors declare that the research was conducted in the absence of any commercial or financial relationships that could be construed as a potential conflict of interest.

\section*{Funding}
YR thanks the CECAM (Centre Europ{\'e}en de Calcul Atomique et Mol{\'e}culaire; Head-director: I. Pagonabarraga) for financial support. DL acknowledges Ministerio de Ciencia, Innovaci\'on y Universidades MCIU/AEI/FEDER for financial support under grant agreement RTI2018-099032-J-I00. 

\section*{Acknowledgments}
We warmly thank Elisabeth Agoritsas for helpful discussions. 

\section*{Data Availability Statement}
The Julia code generating the data used for this study can be found on GitHub: \textcolor{blue}{\href{https://github.com/yrouzaire/forced_xy}{yrouzaire/forced\_xy}}.

\bibliographystyle{Frontiers-Vancouver} 
\bibliography{biblio.bib}



\end{document}